\documentclass[ reprint,
superscriptaddress,
amsmath,
amssymb,
 aps,
 prl
  ]{revtex4-1}
\usepackage{bm}
\usepackage{amsmath}
\usepackage{amssymb}
\usepackage{textcase}
\usepackage{graphicx}
\usepackage{hhline,float}
\usepackage{url}
\usepackage{hyperref}
\newcommand{\unit}[1]{\ensuremath{\,\mathrm{#1}}}
\begin{document}
\title{Measuring nanomechanical motion with an imprecision far below the standard quantum limit}
\author{G.\ Anetsberger$^{1}$, E.\ Gavartin$^{2}$, O.\ Arcizet$^{1}$, 
Q.\ P.\ Unterreithmeier$^{3}$,\\ E.\ M.\ Weig$^{3}$, M.\ L.\ Gorodetsky$^{4}$, J.\ P.\ Kotthaus$^{3}$, T.\ J.\ Kippenberg$^{2,}$}
\email{tobias.kippenberg@epfl.ch}
\affiliation{Max-Planck-Institut f{\"u}r Quantenoptik, Hans-Kopfermann-Str.\ 1, 85748 Garching, Germany\\
$^{2}$ Ecole Polytechnique F$\acute{e}$d$\acute{e}$rale de Lausanne, EPFL, 1015 Lausanne, Switzerland.\\
$^{3}$ Fakult{\"a}t f{\"u}r Physik and Center for NanoScience (CeNS), Ludwig-Maximilians-Universit{\"a}t (LMU),
Geschwister-Scholl-Platz 1, 80539 M{\"unchen}, Germany\\
$^{4}$ Department of Physics, Moscow State University, Moscow 119899, Russia.}

{\abstract
We demonstrate a transducer of nanomechanical motion based on cavity enhanced optical near-fields capable of achieving a shot-noise limited imprecision more than $10 \,\mathrm{ dB}$ below the standard quantum limit (SQL). Residual background due to fundamental thermodynamical frequency fluctuations allows a total imprecision $3 \,\mathrm{ dB}$ below the SQL at room temperature (corresponding to $(600 \unit{am/\sqrt{Hz}})^2$ in absolute units) and is known to reduce to negligible values for moderate cryogenic temperatures. The transducer operates deeply in the quantum backaction dominated regime, prerequisite for exploring quantum backaction, measurement-induced squeezing and accessing sub-SQL sensitivity using backaction evading techniques.
}
\maketitle
The development of nanomechanical oscillators \cite{Ekinci05,*CraigheadScience00} has enabled experiments ranging from precision measurements \cite{Cleland98,*Jensen08,*Rugar04} to studies approaching the quantum regime of mechanical oscillators \cite{SchwabPT05,LaHaye04,Teufel08}. All these experiments rely on the sensitive detection of nanomechanical motion and during the last decades a variety of transducers has been developed. A natural scale for comparison is their ability to achieve a measurement imprecision sufficient to resolve a mechanical oscillator's zero-point motion, which coincides with the sensitivity at the standard quantum limit (SQL) of continuous position measurement \cite{Braginsky74,*Braginsky78,Caves81}. At the SQL the sensitivity of a displacement measurement $S_{xx}^\mathrm{SQL}$ is equal to the quantum mechanical zero-point fluctuations (ZPF) of the nanomechanical oscillator under scrutiny, i.e. $S_{xx}^\mathrm{SQL}=S_{xx}^\mathrm{zpf}$. At resonance these are given by $S_{xx}^{\mathrm{zpf}}=2 \hbar/(m \Omega_\mathrm{m}\Gamma_\mathrm{m})$, where $m$, $\Gamma_\mathrm{m}/2\pi$ and $\Omega_\mathrm{m}/2\pi$ denote effective mass, damping rate and resonance frequency of the nanomechanical oscillator (note that single-sided spectra are used throughout this work).

\begin{figure}[b!]
\centering\includegraphics[width=3in]{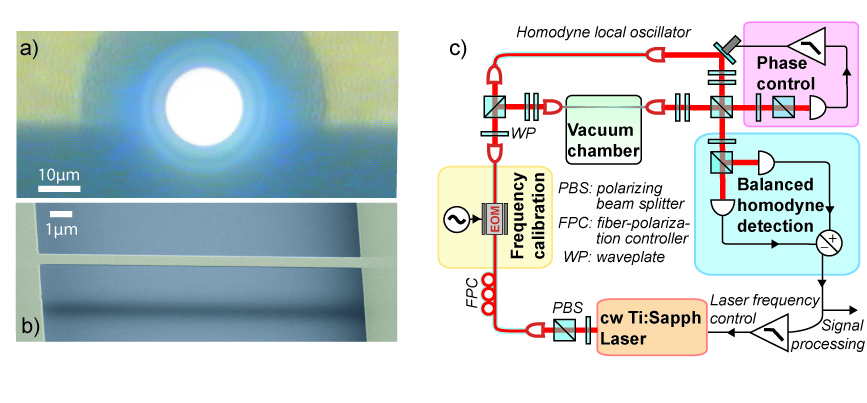}
\caption{Optical and scanning electron micrographs of a $16  \unit {\mu m}$ radius microresonator (a) and a $100\unit{nm} \times 500\unit{nm}\times 15 \unit{\mu m}$ strained SiN nanomechanical oscillator (b). (c) Schematic of the quantum-limited optical homodyne detection scheme. The samples are mounted in a vacuum chamber. A cw titanium sapphire laser emitting at $\lambda =850 \unit{nm}$ is fibre-coupled and locked to cavity resonance using a balanced homodyne interferometer.}
\label{f:1}
\end{figure}
Although in principle the measurement imprecision can be made arbitrarily small by increased coupling strengths between transducer and measured obkect, reaching an imprecision at the level of the ZPF (prerequisite for reaching SQL sensitivity) remained an elusive goal for decades owing to both experimentally limited coupling strengths and excess classical noise. The lowest imprecision has so far been achieved by transducers of nanomechanical motion based on electronic current flow using single-electron transistors \cite{Knobel03,LaHaye04}, SQUIDs \cite{Etaki08} or quantum point contacts \cite{Flowers07,*Poggio08} and has approached \cite{Knobel03,LaHaye04,Regal08,Etaki08,Flowers07,*Poggio08} but not reached the SQL. Only recently, the application of optical near-fields and the development of a near quantum-limited microwave amplifier \cite{Castellanos08} have allowed an imprecision at \cite{Anetsberger09} and slightly below the SQL \cite{Teufel09}. 
Here we demonstrate a transducer of nanomechanical motion with a quantum (shot-noise) limited imprecision $11 \unit{dB}$ below the SQL, i.e. $S_{xx}=0.08 \times S_{xx}^\mathrm{SQL}$. The transducer allows operating deeply in the quantum backaction (QBA) dominated regime with coupling strengths exceeding those required to reach the SQL by more than two orders of magnitude. At room temperature, the total imprecision reaches values $3 \unit{dB}$ below the SQL, limited by fundamental thermorefractive frequency noise. Its magnitude is in excellent agreement with theory and expected to decrease by $\geq 25 \unit{dB}$ for operation below $30 \unit{K}$, i.e. to a negligible level. The presented approach thereby paves the way to studying the QBA of radiation pressure \cite{Caves81,Verlot09} that ultimately enforces the SQL but has never been observed to date. Operating in the regime where QBA should be the dominating contribution to sensitivity may moreover enable measurement induced squeezing \cite{Fabre94} and leverage the use of backaction evading techniques \cite{BraginskyB92,*Hertzberg10}.

Fig.\ \ref{f:1} shows the experimental setup. A small mode volume and high optical $Q$ ($Q>10^8$) toroid microresonator \cite{KippenbergAPL04} (Fig.\ \ref{f:1}a) is fibre-coupled using a fibre-taper. The strained SiN nanomechanical oscillator \cite{Verbridge06,*Unter09} (Fig.\ \ref{f:1}b, typical eigenfrequencies $\Omega_\mathrm{m}/2\pi\approx 10 \unit{MHz}$) interacts with the microresonator by virtue of the evanescent field \cite{Anetsberger09,Braginsky83} of its optical whispering gallery modes (WGM). The interaction is described by two parameters: the dispersive optomechanical coupling coefficient $g=d\omega/dx$  and the reactive contribution $\gamma=d\kappa/dx$ ($\omega/2\pi$: cavity resonance frequency, $\kappa/2\pi$: cavity energy decay rate). Both depend on the separation $x$ between the microresonator and the nanomechanical oscillator which is controlled using piezoelectric positioners. Coupling coefficients $g/2\pi >50 \unit{MHz/nm}$ can be achieved which corresponds to a frequency shift of the optical resonance (wavelength $ \lambda=850 \unit{nm}$) by more than $10 \unit{GHz}$ and represents a more than tenfold improvement compared to previous work \cite{Anetsberger09}. At the same time, the cavity linewidth is only slightly affected. We note that we have achieved coupling rates larger than $200\unit{MHz/nm}$, at the expense of increased reactive coupling rates ($\gamma/2\pi \approx 10 \unit{MHz/nm}$). All measurements shown in this letter are, however, performed at $\gamma/2\pi \leq 0.2 \unit{MHz/nm} \ll g/2\pi$.

The near-field interaction thus dispersively transduces the nanomechanical oscillator's Brownian noise displacement spectrum $S_{xx}^\mathrm{n}[\Omega]$ into frequency noise of the microresonator $S_{\omega\omega}^\mathrm{n}[\Omega]=g^2\times S_{xx}^\mathrm{n}[\Omega]$. Our measurement system 
\begin{figure}[t!]
\centering\includegraphics{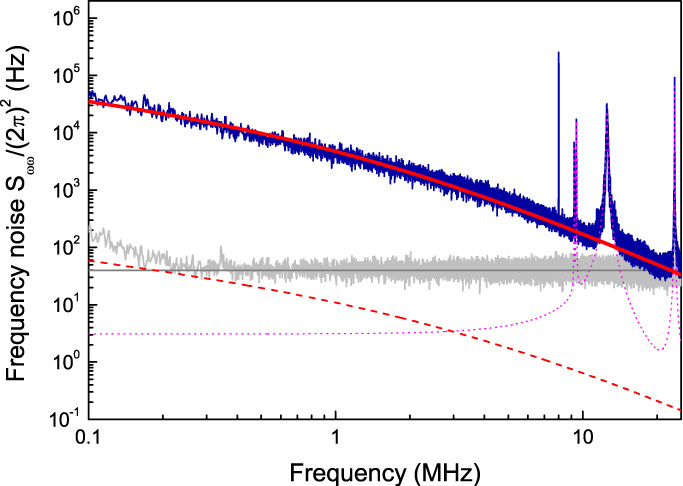}
\caption{Frequency noise background of the microresonator. The predominant broadband noise is due to thermorefractive frequency noise (red line) fitted according to Eq. (1). For a bath temperature of $30 \unit{K}$, a $25 \,\mathrm{dB}$ reduction of thermorefractive noise is expected (red dashed line). A fit to the intrinsic mechanical modes of the optical microresonator (pink dashed line) shows that their off-resonant contribution is negligible (peak at $8 \unit{MHz}$: frequency calibration marker). The equivalent frequency noise caused by shot-noise (grey) lies far below the thermorefractive noise (at $P_\mathrm{in}\sim 10 \unit{\mu W}$).}
\label{f:2}
\end{figure}
consists of a shot-noise limited (for Fourier frequencies $\Omega> 500 \unit{kHz}$), continuous wave (cw) titanium sapphire laser combined with a (near unity efficiency) balanced homodyne interferometer, as shown in Fig.\ \ref{f:1}c. The balanced homodyne detection enables quantum (shot-noise) limited measurements of the microresonator's frequency noise even for very weak probe powers ($\sim \mathrm{\mu W}$). We employ two individual locks balancing the phase difference of the interferometer arms and locking the laser strictly to line-centre of a microresonator mode independently. Thus, dynamical backaction due to non-zero laser detuning is avoided which could lead to linewidth-narrowing and artificially increased displacement spectral densities of the nanomechanical oscillators \cite{Anetsberger09}. This is particularly important, since the optomechanical interaction is strong enough to reach the threshold for the parametric instability \cite{Braginsky01,*Kippenberg05} for only $400 \unit{nW}$ of input power.

Fig.\ \ref{f:2} shows a spectrum of the frequency noise of a microresonator (radius $R= 18 \unit{\mu m}$) alone, without coupling to the nanomechanical oscillator. Owing to the quantum-limited detection scheme the background of the measurement is only given by laser shot-noise \cite{Anetsberger09}
\begin{eqnarray}
S_{\omega\omega}^\mathrm{shot}[\Omega] =\frac{\hbar \omega}{P_\mathrm{in}} \frac{\kappa^2}{8} \left(1+4 \Omega^2/\kappa^2\right)\,.
\label{eq:shot}
\end{eqnarray}
By using sufficiently large input power ($P_\mathrm{in}\gtrsim 5\unit{\mu W}$), its magnitude can be lowered far below both other contributing frequency noise sources:
First, the contribution $S_{xx}^\mathrm{\mu}[\Omega]$ of intrinsic mechanical modes of the microresonator itself \cite{SchliesserNJP08} leads to frequency noise according to $S_{\omega\omega}^\mathrm{\mu}[\Omega]=(\omega/R)^2\times S_{xx}^\mathrm{\mu}[\Omega]$. The frequency and mechanical quality factor of these modes can be optimized via the resonator geometry \cite{Anetsberger08,SchliesserNJP08} and allows confining this source of frequency noise to narrow frequency bands. The second contribution is broadband in nature and given by the fundamental temperature fluctuations  $S_{TT}[\Omega]$ within the cavity mode volume \cite{Gorodetsky04} which lead to thermorefractive cavity frequency noise $S_{\omega\omega}^\mathrm{thr}[\Omega]=(\omega/n \cdot dn/dT)^2\times S_{TT}[\Omega]$ ($n$: refractive index of silica). For a fundamental WGM of a microtoroid it is given by: 
\begin{eqnarray}
S_{\omega\omega}^{\mathrm{thr}}[\Omega]=\left(\frac{\omega}{n} \frac{dn}{dT}\right)^2  \frac{(16 \pi)^{\frac{1}{3}} k_B T^2}{V \rho C }\frac{\sqrt{\tau/\Omega}}{(1+(\Omega \tau)^{\frac{3}{4}})^2}\, ,
\label{eq:1}
\end{eqnarray}
where $C$ and $\rho$ are the heat capacity and density of silica. $V$ and $\ell$ denote the mode volume and the angular mode number of the optical mode, approximated by a gaussian ellipse with semi-axes $b=0.77 R/\ell^{2/3}$ and $d=R^{3/4}r^{1/4}/\ell^{1/2}$ ($r$: minor radius of the toroid). The thermal cut-off time $\tau$ is given by  $\tau^{-1}=(4/\pi)^{1/3} D\, (b^{-2}+d^{-2})$, where $D$ denotes the thermal diffusivity of silica. In order to fit Eq.\ (\ref{eq:1}) to 
the measured data, the semi-axes $b$ and $d$ are used as fit parameters. Excellent agreement as shown in Fig.\ \ref{f:2} is found with $b$ ($d$) decreased (increased) by $\approx 25\%$ and all other parameters being fixed. Thus, this source of frequency noise is extremely well understood.

In the following we place a SiN nanomechanical oscillator in the optical near-field.
The total measured displacement spectrum $S_{xx}[\Omega]$ thus consists of the following contributions: 
\begin{eqnarray}
&S_{xx}[\Omega] =& 
S_{xx}^\mathrm{n}[\Omega] \nonumber\\
&\hphantom{S_{xx}[\Omega]}& +\frac{1}{g^2} \left(
S_{\omega\omega}^\mathrm{shot}[\Omega]+S_{\omega\omega}^\mathrm{thr}[\Omega]+\frac{\omega^2}{R^2} S_{xx}^\mathrm{\mu}[\Omega]
\right)
\,.
\label{eq:3}
\end{eqnarray}
\begin{figure}[t!]
\centering\includegraphics{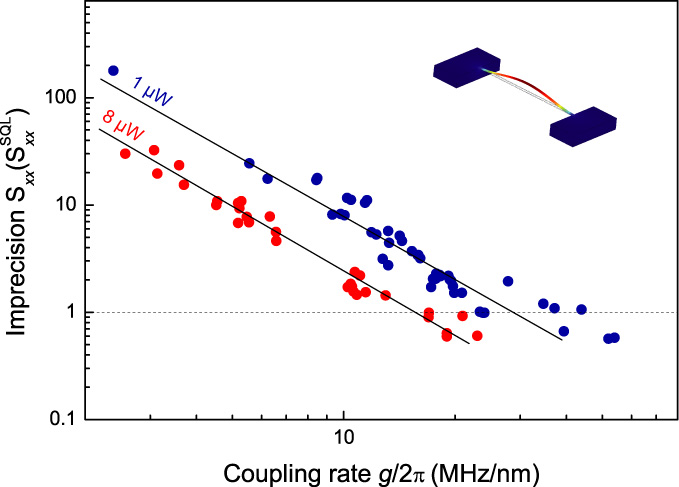}
\caption{Measurement of a high frequency nanomechanical oscillator with an imprecision below $ S_{xx}^\mathrm{SQL}$. For two different input powers $1 \unit{\mu W}$ (blue) and $8 \unit{\mu W}$ (red) the absolute measurement imprecision is shown to reduce with increased optomechanical coupling rate $g/2\pi$ according to Eq.\ (\ref{eq:3}). Both power levels allow reaching an imprecision below the SQL, achieving values of $S_{xx}=0.5\times S_{xx}^\mathrm{SQL}$ for a $8.3 \unit{MHz}$ nanomechanical oscillator ($30\unit{\mu m}\times 700 \unit{nm}\times 100\unit{nm}$, $m=3.7\cdot10^{-15}kg$). Inset: finite element simulation of its fundamental mechanical mode.}
\label{f:3}
\end{figure}
It can be clearly seen that increasing the optomechanical coupling $g$ (that is 
reducing the distance between the nanomechanical oscillator and the cavity) equally reduces all background noises, independent of their origin. This is shown in Fig.\ \ref{f:3} where the measurement imprecision (i.e. the full background of the measurement) obtained for a nanomechanical string is depicted as a function of the optomechanical coupling rate. Increasing the optomechanical coupling rate allows obtaining an imprecision $3 \unit{dB}$ below the SQL both for $1 \unit{\mu W}$ and $8 \unit{\mu W}$ input powers. This represents the first optical measurements of nanomechanical motion with an imprecision below the SQL. A closer look at the different background noise contributions reveals that the employed power levels are already far beyond the power level needed to reach the SQL ($P_\mathrm{SQL}$) thus falling deeply in the QBA dominated regime. In an ideal measurement with a lossless cavity and in the absence of thermal photons $P_\mathrm{SQL}$ is given by
\begin{eqnarray}
P_\mathrm{SQL}/\hbar\omega=\frac{(\kappa/4)^2}{g^2 S_{xx}^\mathrm{zpf}}\left(1+\frac{4 \Omega_\mathrm{m}^2}{\kappa^2}\right)\,.
\label{eq:2}
\end{eqnarray}
Eq.\ (\ref{eq:2}) reveals that a photon flux of one photon per second will be sufficient to reach the SQL if the zero-point motion of the mechanical oscillator (across a bandwidth of $1 \unit{Hz}$) 
shifts the cavity resonance by more than a quarter of its linewidth. In our measurements we use a critically coupled (impedance matched) cavity with a total linewidth $\kappa$ given to equal parts by outcoupling $\kappa_\mathrm{ex}$ and intrinsic losses ($\kappa_\mathrm{0}= \kappa_\mathrm{ex}$). Therefore, the power to reach the SQL is slightly increased to $P_\mathrm{SQL}^\mathrm{cc}=\sqrt{8}\times P_\mathrm{SQL}$. Since experimentally available power levels are limited and high power levels in addition enhance unwanted thermal effects, minimizing $P_\mathrm{SQL}$ is generally desirable. Interestingly, the threshold power for the parametric oscillation instability \cite{Braginsky01,*Kippenberg05} $P_\mathrm{th}$ or alternatively the power needed to cool the mechanical oscillator by a factor of two scales similarly: $P_\mathrm{th}=(4+\kappa^2/4 \Omega_\mathrm{m}^2) \times P_\mathrm{SQL}$, with $P_\mathrm{th} \cong 4\times P_\mathrm{SQL}$ in the resolved sideband regime \cite{SchliesserNP08}. Therefore, $P_\mathrm{SQL}$ is a figure of merit describing both a transducer's ability to manipulate the mechanical oscillator via dynamical backaction \begin{figure}[t!]
\centering\includegraphics{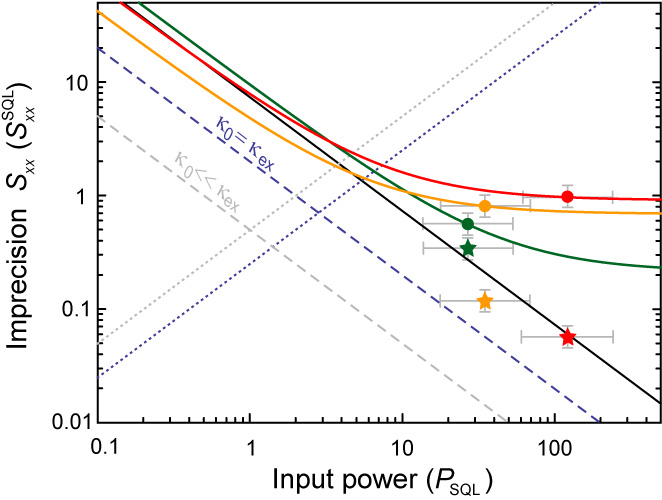}
\caption{Shot-noise limited imprecision more than $10 \unit{dB}$ below the SQL. The measurement imprecision is shown versus optical input power. The dashed (dotted) lines denote the shot-noise (QBA) contribution for an ideal, lossless measurement (grey) and for an impedance matched cavity (blue). The solid lines show the full imprecision for three measurements (dots) with different levels of thermorefractive noise (owing to different optomechanical coupling rates). With increasing input power levels, the measured shot-noise limited imprecision (stars) lies far below the SQL, reaching a lowest value of $0.08\times S_{xx}^\mathrm{SQL}$, for $P_\mathrm{in}=122\times P_\mathrm{SQL}$, thus falling deeply in the QBA dominated regime. The black line is a fit to the measured shot-noise contribution.}
\label{f:4}
\end{figure}
and to reach the SQL.

\begin{figure}[t!]
\centering\includegraphics{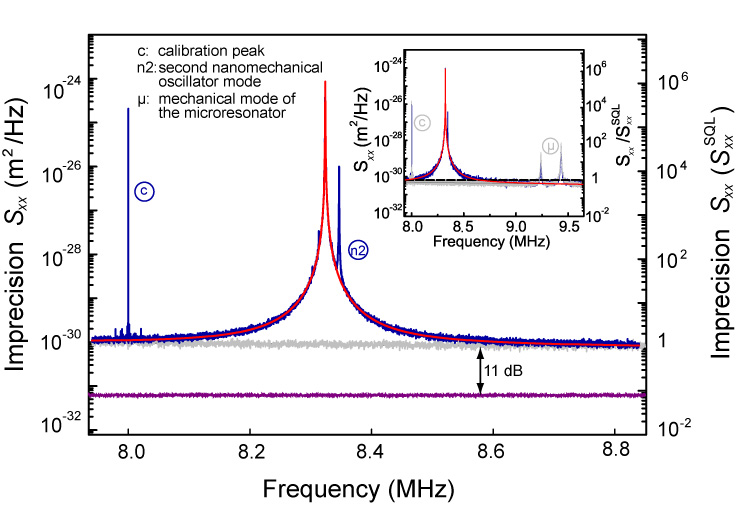}
\caption{The main panel shows the spectrum corresponding to the red data in Fig.\ \ref{f:4}. The full measurement imprecision (grey) is slightly below the SQL. The shot-noise level, however, lies $11 \unit{dB}$ below the SQL at $S_{xx}=0.08\times S_{xx}^\mathrm{SQL}$ or $(250 \unit{am/\sqrt{Hz}})^2$ in absolute units. Inset: trace corresponding to the green data in Fig.\ \ref{f:4} with lower input power but higher optomechanical coupling and a total measurement imprecision of $0.5\times S_{xx}^\mathrm{SQL}$ or $(600 \unit{am/\sqrt{Hz}})^2$.}
\label{f:5}
\end{figure}

Fig.\ \ref{f:4} shows the ideal imprecision as well as the measured shot-noise limited imprecision (stars) and total imprecision with a given level of thermorefractive noise (dots) for three different input power levels. Already an input power of $P_\mathrm{in}=1 \unit{ \mu W}$ corresponds to $P_\mathrm{in}=27 \times P_\mathrm{SQL}$ and the total imprecision which is dominated by shot-noise reaches a value of $S_{xx}=0.5\times S_{xx}^\mathrm{SQL}$. These data correspond to the trace shown in Fig.\ \ref{f:5} (inset) recorded with a coupling rate of $50 \unit{MHz/nm}$. As shown in Fig.\ \ref{f:4}, increasing the input power reduces the shot-noise contribution to the imprecision further and the measurement background given by shot-noise can be lowered to values far below the SQL. Using an input power of $P_\mathrm{in}=8 \unit{\mu W}$ (corresponding to $P_\mathrm{in}=122\times P_\mathrm{SQL}$) allows reaching a shot-noise limited imprecision of $0.08\times S_{xx}^\mathrm{SQL}$, i.e. $11 \unit{dB}$ below the SQL, or $S_{xx}^\mathrm{shot}= (250 \unit{am/\sqrt{Hz}})^2$ in absolute units. Fig.\ \ref{f:5} shows the corresponding spectrum recorded with a coupling rate of $17 \unit{MHz/nm}$. Thus, for input powers of only a few micro-Watts and despite the comparatively low mechanical quality factor ($30'000$) the measurements are 
already in a regime where the shot-noise limited imprecision reaches values far below the SQL whereas the dominating contribution to sensitivity should be given by QBA, which is expected to be more than two orders of magnitude larger than the shot-noise contribution (see Fig.\ \ref{f:4}). At present, this contribution is masked by the Brownian motion of the nanomechanical oscillator. By using lower frequency and higher $Q$ mechanical oscillators, the QBA could, however, be increased to values comparable to the thermal noise even at room temperature \cite{Anetsberger09}.

Owing to the close connection between dynamical backaction and $P_\mathrm{SQL}$, small imperfections in the laser detuning can lead to strong dynamical backaction effects. Reliably ruling these out therefore requires slightly reducing the optomechanical coupling rates when working with increased power levels. Consequently, the higher power measurements are partly limited by thermorefractive noise (which as opposed to shot-noise does not reduce for higher laser power). The thermorefractive noise background ($S_{\omega\omega}^{\mathrm{thr}}[\Omega]  = (2\pi \, 14 \unit{\sqrt{Hz}})^2$ at $\Omega/2\pi\sim 8.5\unit{MHz}$) however still allows a total imprecision $3 \unit{dB}$ below the SQL at all power levels ($1$--$10 \unit{\mu W}$) employed in this work as shown in Fig.\ \ref{f:3}. We note that with the well-known material parameters of silica, the thermorefractive noise according to Eq.\ (\ref{eq:1}) is suppressed by $\geq 25 \unit{dB}$ when operating at $\leq 30 \unit{K}$ instead of ambient temperature (see Fig.\ \ref{f:2}). Thus, thermorefractive noise will be negligible at low temperatures (without even taking into account the large improvements in mechanical $Q$ at low temperatures that increase the SQL). This represents a major advantage compared to even the most sensitive transducers based on microwaves which are currently also limited by cavity frequency noise albeit operating at temperatures of $130 \unit{mK}$ \cite{Teufel09}.

Moreover, our measurement scheme offers both large flexibility concerning oscillator material and geometry, and high bandwidth ($\sim 100 \unit{MHz}$). The sampled length of the nanomechanical oscillator, approximately given by $\sqrt{\lambda R/2}$, amounts to only $\sim 2.5\unit{\mu m}$ for $\lambda \sim 850 \unit{nm}$ and cavity radii of $15-20 \unit{\mu m}$. Thus, the transducer can also be applied to shorter, higher frequency nanomechanical oscillators. With the parameters used above ($g/2\pi =50\unit{ MHz/nm}$, $\kappa/2\pi =60 \unit{MHz}$), e.g. for a beam of $5 \unit{\mu m}$ length ($\Omega/2\pi=50 \unit{MHz}$, $m=5 \times 10^{-16}\unit{kg}$, $Q=50'000$) the SQL could be reached for $\cong 1\unit{\mu W}$.

T.J.K. acknowledges financial support by an ERC Starting Grant (SiMP), MINOS, a Marie Curie Excellence Grant and continued support by MPQ. J.P.K. acknowledges financial support by the Deutsche Forschungsgemeinschaft (Ko 416/18), LMUexcellent, and LMUinnovativ. T.J.K., J.P.K., E.M.W. acknowledge the German Excellence Initiative via the Nanosystems Initiative Munich (NIM). M.L.G. acknowledges the Alexander von Humboldt foundation and the Dynasty Foundation support programme.

\end{document}